\documentclass[onecolumn,authoryear]{els-mrw}

\usepackage{amsmath,amssymb,amsfonts,amsthm,makeidx,graphicx}
\usepackage{txfonts}
\usepackage{helvet}
\usepackage{aas_macros}

\begin{document}

\chapter{Main-sequence exoplanet systems: tidal evolution}\label{chap1}

\author[1]{Kaloyan Penev}%

\address[1]{\orgname{The University of Texas at Dallas}, \orgdiv{Department of
Physics}, \orgaddress{Richardson, TX}}

\articletag{
    Chapter Article tagline: update of previous edition,, reprint..
}

\maketitle

\begin{glossary}[Key Points]

    \term{Hot Jupiter:} is an extrasolar planet composed primarily of hydrogen
    and helium (similar to Jupiter and Saturn in our own Solar System) residing
    very close to its parent star.

    \term{Tidal bulge:} is the stretching of an astrophysical object (planet or
    star), approximately in the direction of a nearby massive companion.

    \term{Tidal dissipation:} is the conversion of mechanical energy to heat as
    a tidal bulge moves through a planet or star.

    \term{Tidal lag:} is the offset between the tidal bulge and the line
    connecting the centers of the two objects produced by the tidal dissipation.

    \term{Tidal evolution:} is the change in the orbit and spins of the objects
    in an exoplanet system due to the gravitational coupling between the tidal
    bulges on each object with the other object in the system.

    \term{Synchronous rotation:} is a state of a planet or star in which the
    object's rotational period is equal to the orbital period.

    \term{Pseudo-synchronous rotation:} is the rotation of a planet or star with
    such a period that the orbit averaged tidal torque is zero.

    \term{Tidal circularization:} is one of the effects of tidal dissipation
    causing the orbit of a planet to become more and more circular over time.

    \term{Tidal inspiral:} is the shrinking of the orbit of a planet due to
    tides, bringing it closer and closer to its parent star.

    \term{Tidal alignment:} is a process which gradually aligns the spin angular
    momenta of the star and the planet to the orbital angular momentum.

    \term{Tidal heating:} refers to the heat deposited within a planet by tidal
    dissipation. This heat can be a non-negligible contribution to the energy
    budget of the planet and can drive various processes.

\end{glossary}

\begin{abstract}[Abstract]

The easiest exoplanets to detect are those that orbit very close to their host
stars. As a result, even though these planets are quite rare, they represent a
major fraction of the current exoplanet population. A side-effect of the
proximity between the planet and the star is that the two have strong mutual
interactions through a number of physical processes. One of the most important
of these processes is tides. Tides are thought to shape the orbits of close-in
exoplanets, heat the planet making its radius expand, and even drive some
planets to spiral into their host stars. This chapter briefly introduces the
basics of tidal physics and describes the various fingerprints tides leave
within the observed exoplanet population.

\end{abstract}

\section{Introduction}
\label{sec:introduction}

The term tides refers to changes in the shape of an astronomical object in
response to differences in the gravitational pull by a nearby mass at different
locations within the object. In this chapter, tides and their effects are
discussed mostly qualitatively. For a full mathematical description of tides see
\cite{Murray_Dermott_book}.

Consider an exoplanet system consisting of a single planet orbiting a single
star in a circular orbit. The gravitational acceleration due to the star at the
location of the planet's center provides the centripetal acceleration needed to
keep the planet in orbit. The part of the planet facing the star is closer to
the star and therefore experiences a slightly stronger gravitational pull.
However, if we ignore the rotation of the planet for a moment, that part of the
planet must follow a path with the same radius and period as the center of the
planet (though with its center offset slightly). The result is that the star
provides larger gravitational acceleration to that part of the planet than the
required centripetal acceleration. This excess force is referred to as the tidal
force, and it is directed towards the star for the star-facing part of the
planet. Similarly, on the far side of the planet, the star's gravity is slightly
smaller than what is required to keep that part of the planet in orbit,
resulting in a tidal force pointing away from the star. Relative to the planet
center, both of these forces point outward. Forces internal to the planet (e.g.
self gravity) must balance those extra pulls, causing the planet to elongate
along the star-planet line, and squeeze in the perpendicular direction. This
elongation is frequently referred to as the tidal bulge.

Let us now consider the rotation of the planet. Similarly to tides, rotation
slightly decreases the outward force near the equator of the planet required to
counteract gravity, since part of the gravitational force is required to provide
centripetal acceleration. This causes a slight equatorial bulge. In addition,
rotation interacts with tides. If the period of rotation is exactly equal to the
orbital period (a.k.a. synchronous rotation), the sub-stellar point is fixed on
the planet's surface, and consequently the tidal bulge is also fixed relative to
the planet. However, if the rotation and orbital periods differ, the tidal bulge
will travel on the planet's surface. If the rotational angular velocity of the
planet is $\Omega_{pl}$, and the orbital angular velocity is $\Omega_{orb}$,
the sub-stellar point travel will travel on the surface with an angular
frequency $\Omega_{orb} - \Omega_{pl}$. Since there are two tidal bulges on the
planet (one on the side facing the star and one on the opposite side), the
planetary material will experience a tidal wave with a frequency of
$2(\Omega_{orb} - \Omega_{pl})$.

\begin{figure}[t]
    \centering
    \includegraphics[width=0.5\textwidth]{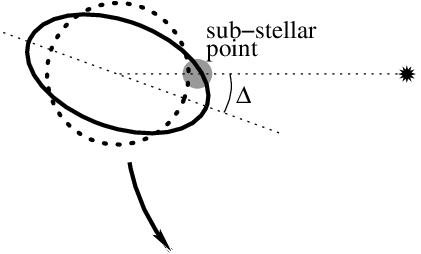}
    \caption{
        Exaggerated tidal bulge on a planet orbiting a star. Assuming that the
        planet spin angular velocity is smaller than than the orbital angular
        velocity, the tidal bulge will lag behind the sub-stellar point by an
        angle $\Delta$.
    }
    \label{fig:tidal_bulge}
\end{figure}

Any time dependent deformation of the planet will result in some conversion of
mechanical energy to heat. The simplest picture of this is friction between
parts of the planet that move relative to each other. For a continuous medium,
this friction is encapsulated in the concept of viscosity. However, a great
variety of physical processes can result in energy dissipation of tidal
perturbations. The physical causes and the amount of energy that is dissipated
can in principle depend on the internal structure and spin of the planet, as
well as on the frequency and amplitude of the tidal forcing. Regardless of the
causes, this energy dissipation introduces a delay between the tidal forcing and
the response of the planet to that forcing. If the planet spins faster than
synchronous (i.e.  $\Omega_{pl} > \Omega_{orb}$), the tidal bulge will be
carried ahead of the sub-stellar point by the planet's rotation, and if
$\Omega_{pl} < \Omega_{orb}$, the tidal bulge will lag behind
(Fig.~\ref{fig:tidal_bulge}).

The two tidal bulges, now shifted relative to the star-planet line, will
experience the gravitational pull of the star. Since the closer bulge will feel
a stronger gravitational force than the farther bulge, there will be a net
torque on the planet. If the bulge is carried ahead of the sub-stellar point by
rotation, the gravitational pull of the star will apply a torque to the planet
opposite to its rotation, acting to slow down its spin, and the reaction force
on the star will act to add angular momentum to the orbit. Conversely, if the
tidal bulge lags behind the sub-stellar point, the gravitational pull of the
star will act to spin the planet up, taking angular momentum out of the orbit.

The discussion above assumed that the equator of the planet is aligned with the
orbital plane. This is not necessarily the case. If the planet's spin is tilted
with respect to the orbit, regardless of whether the planet is rotating faster
or slower than the orbit, rotation will shift the bulges away from the
sub-stellar point in such a way as to produce a torque that will tend over time
to bring the planet's equator and the orbital plane into alignment.

So far we only considered circular orbits. For non-circular orbits, the orbital
angular velocity is no longer constant. It is highest near periapsis (closest
approach between the planet and star) and lowest near apoapsis (largest
planet-star distance). Because the orbital angular velocity is changing, there
is no rotation period for which the tidal bulge can be static on the surface of
the planet, so no matter the spin of the planet there will always be some
non-zero tidal dissipation. In this case, there exists a spin angular velocity
of the planet, for which the average tidal torque over an orbit vanishes. This
is known as pseudo-synchronous rotation. Near periapsis the orbital angular
velocity exceeds the pseudo-synchronous spin of the planet, thus the tidal bulge
leads the sub-stellar point, causing angular momentum to flow from the orbit to
the planet. Then for the part of the orbit around apoapsis, the orbital angular
velocity is smaller than the pseudo-synchronous spin of the planet, causing
angular momentum to flow in the opposite direction.

\begin{figure}[t]
    \centering
    \includegraphics[width=0.5\textwidth]{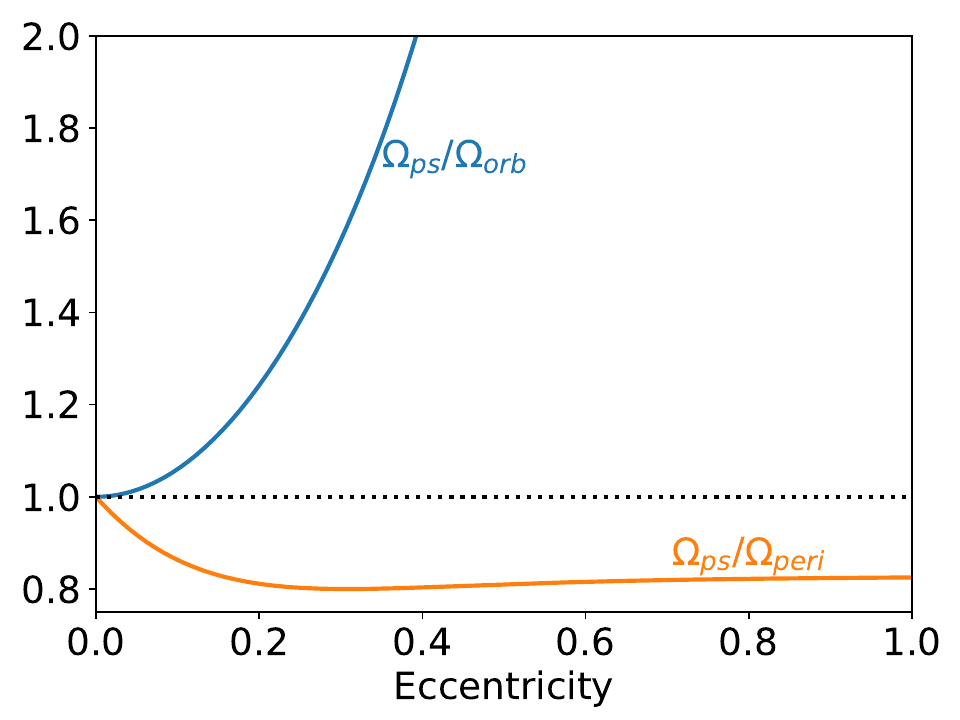}
    \caption{
        The ratio of the pseudo-synchronous spin angular velocity of a planet to
        the average orbital angular velocity and the orbital angular velocity at
        pericenter as a function of the orbital eccentricity assuming the time
        lag of the tidal bulge relative to the tidal forcing is independent of
        the forcing frequency.
    }
    \label{fig:pseudo-synchronous}
\end{figure}

The exact value of the pseudo-synchronous spin depends on the properties of the
tidal dissipation mechanism. \citet{Hut_81} derived an expression for it under
the assumption that all tidal waves experience the same time lag regardless of
the frequency of the tidal forcing giving rise to them. Under that assumption,
since tides are stronger near periapsis, the pseudo-synchronous period must be
somewhat shorter than the orbital period, resulting in the planet being
spun-down over a larger fraction of the orbit than it is being spun up, but at a
slower rate. Thus, the \citet{Hut_81} pseudo-synchronous rotational angular
velocity ($\Omega_{ps}$) of the planet lies between the orbit-average angular
velocity ($\Omega_{orb}$) and the angular velocity at periapsis ($\Omega_{peri}
= \sqrt{(1+e)/(1-e)^3}\Omega_{orb}$). Figure \ref{fig:pseudo-synchronous} shows
the eccentricity dependence of $\Omega_{ps}$ under these assumptions in units of
the orbital angular frequency and the angular frequency at periapsis. If the
spin of the planet is faster(slower) than the pseudo-synchronous rate tides will
act to spin the planet down(up).

For eccentric orbits, even if the planet is pseudo-synchronized, tidal evolution
does not stop. Even though the average torque is zero, energy is still being
dissipated because the sub-stellar point is not fixed on the surface and the
planet-star distance varies over time. From the discussion about
pseudo-synchronous period above, it follows that near periapsis, the sub-stellar
point will drift eastward (i.e. in the direction of rotation), and near apoapsis
it will drift westward, with a long-term westward average, since the
pseudo-synchronous spin period is shorter than the orbital period. These
shifting both in location and amplitude tidal bulges will still experience
energy dissipation. Thus while on average no angular momentum will be exchanged
between the planet and the orbit, energy will still be extracted from the system
and dissipated as heat, causing the orbit to circularize over time.

Everything we have said so far about the tides the star raises on the planet
applies equally to the tides the planet raises on the star. For circular orbits,
aligned with the stellar equator, if the spin angular velocity of the star
exceeds the orbital angular velocity, angular momentum is transferred from the
stellar spin to the orbit, and if the stellar spin is slower than the orbit,
angular momentum flows in the opposite direction. Similarly, we can define a
pseudo-synchronous period for the star at which the net tidal torque averaged
over a complete orbit vanishes. Even if the star is spinning
pseudo-synchronously, tidal energy dissipation continues and acts to circularize
the orbit over time. For misaligned orbits, tides will gradually work to align
the orbital plane with the stellar equator.

The amplitude of the tidal bulge is set by competition between the tidal force
and the self-gravity of the planet or the star experiencing the tides.
Consequently, the planetary tides are positively correlated with the planet
radius and stellar mass and negatively correlated with the planet mass.
Conversely, the stellar tides are larger for more massive planets and smaller
for larger stellar masses. Both stellar and planetary tides get weaker with
increasing star-planet separation, implying that tides will only be important if
the planet gets close to its parent star for at least some part of its orbit.
This can occur in one of two ways: either the planet has a very small orbital
semimajor axis (short period orbit), or the eccentricity of the planet is very
large making the pericenter distance small.

\section{Tidal timescales}
\label{sec:timescales}

Based on the discussion above, we can think of four separate tidal evolution
timescales (ordered from shortest to longest):

\begin{enumerate}
    \item The timescale on which the planet's spin gets pseudo-synchronized and
        aligned with the orbit
    \item The timescale on which the orbit circularizes
    \item The timescale on which the orbital period of a circular orbit changes
    \item The timescale on which the stellar spin changes
\end{enumerate}

All of these timescales are orders of magnitude longer than the orbital period.
If they were not, tides would be strong enough to tear the planet apart. This
allows for a very important simplification when calculating the tidal evolution.
Namely, one can begin by neglecting the evolution of the shape and size of the
orbit, as well as the spins of the two object over a single orbital period.
Under these assumptions, one could calculate the average torque and power due to
tides over an orbital period. The evolution of the orbital elements and stellar
and planetary spins can then be calculated based on these averages.

Let us compare the timescales listed above. We begin by considering the relative
angular momenta of the planet's spin, the orbit, and the stellar spin. The first
scales as the mass of the planet times its radius squared. The second scales as
the mass of the planet times the size of the orbit squared. The third scales as
the mass of the star times the stellar radius squared. Since the planet radius
is tiny compared to the orbit and the planet mass and radius are both very small
compared to the star's, the planet's spin angular momentum is negligible
compared to both the orbital angular momentum and the spin angular momentum of
the star. Consequently, both the direction and period of the planet's spin can
be changed by tides without significantly affecting the orbit, implying that the
timescale on which the planet's spin gets pseudo-synchronized and aligned with
the orbit is negligible compared to the timescales on which the orbit or the
stellar spin evolve. Thus, we can assume that if a given exoplanet orbits close
enough to its star for tides to be important, to a good approximation we can
assume that the planet's spin is aligned with the orbit and pseudo-synchronized.

To compare the two timescales on which the orbit changes (second and third
timescales above), we begin by comparing the importance of the planetary to the
stellar tides. The tidal deformation of an object is set by a competition
between the tidal force due to the companion and the self-gravity of the object
being tidally stretched. Since the planet has a smaller self-gravity and
experiences much larger tidal force than the star, the planet is much more
tidally deformed than the star. Consequently, as long as planetary tides are not
static they will drive faster orbital evolution than the stellar tides. By the
reasoning above, for both circular and eccentric orbits, we expect the planet to
be aligned and spin pseudo-synchronously with the orbit.  Hence, for eccentric
orbits, the planetary tides still contribute to the evolution, while for
circular orbits they do not. This in turn means that the tidal circularization
timescale is shorter than the timescale on which circular orbits change their
period, since the latter is only driven by the much weaker stellar tides.

To compare these timescales to the timescale on which the stellar spin is
affected by tides, note that stars have sufficiently large moment of inertia
for their spin angular momentum to be comparable to or larger than the orbital
angular momentum. Since only the stellar tides can affect the stellar spin, this
implies that the timescale on which the stellar spin changes is comparable to or
longer than the timescale on which the orbital period changes.

\begin{figure}[t]
    \centering
    \includegraphics[width=0.49\textwidth]{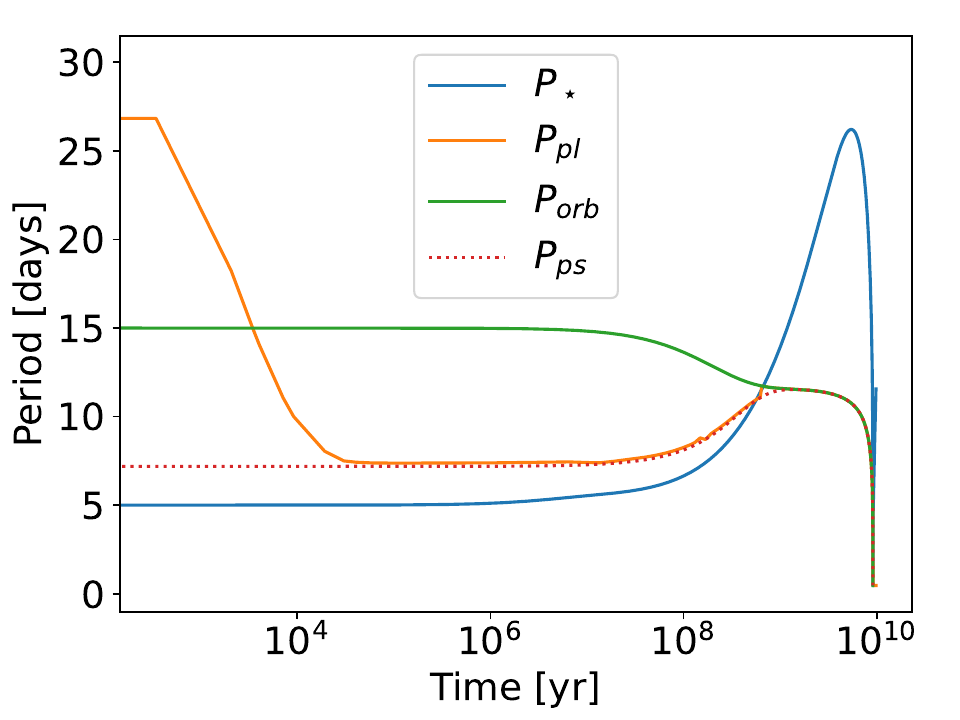}
    \includegraphics[width=0.49\textwidth]{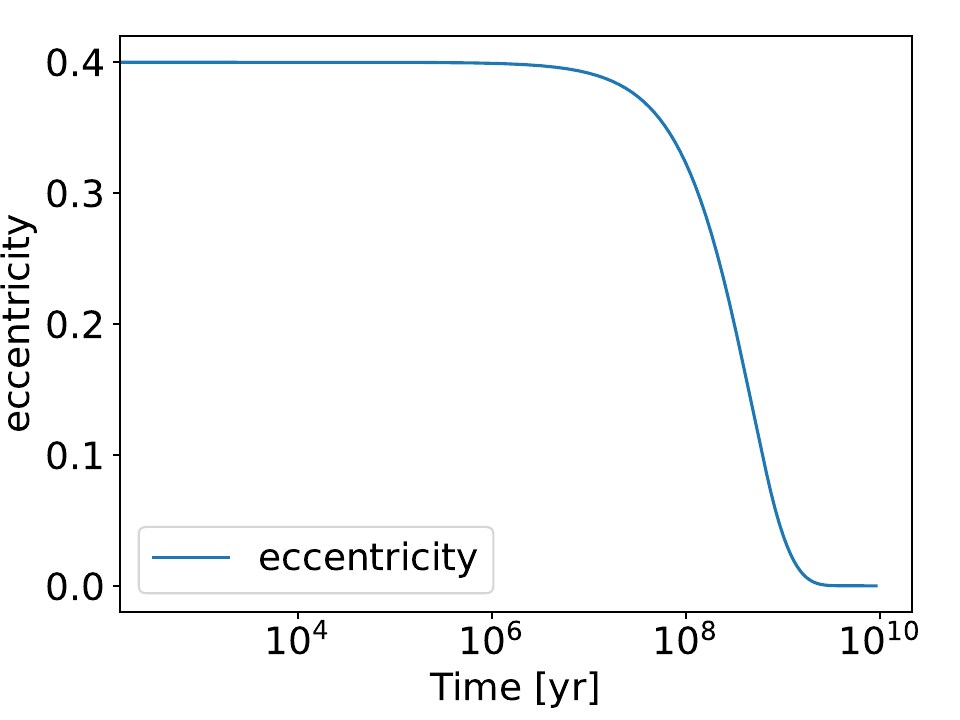}
    \caption{
        Example main-sequence tidal evolution of a Jupiter-mass planet around a
        Solar mass star. The left panel shows the evolution of the orbital
        period (green line) and the spin periods of the planet (orange line) and
        the star (blue line). The dotted red line also shows the
        pseudo-synchronous rotation period for the planet following
        \citet{Hut_81} (i.e. the spin at which the orbit averaged tidal torque
        vanishes). The right panel shows the evolution of the eccentricity.
    }
    \label{fig:example_evolution}
\end{figure}

Figure \ref{fig:example_evolution} shows a theoretically calculated tidal
evolution of a Jupiter-mass planet around a Solar mass star assuming the
simplest possible prescription of tidal friction. Namely, tidal friction is
assumed to produce the same time lag for all tidal waves regardless of their
frequency or spin of the object. The lag was assumed the same for the planet and
the star. Only the main-sequence part of the evolution is shown to highlight the
effect of tides, excluding  earlier ages when the stellar spin evolution is
dominated by the changes in the moment of inertia of the star as it contracts.
The initial configuration of the system at the start of the simulation was
chosen such that all the effects described above occur within the main sequence
lifetime of the star. If the initial orbital period is increased or decreased
all timescales get longer or shorter respectively with a very steep dependence.

Since the time lag and initial conditions for this calculation were somewhat
arbitrarily chosen (though not unreasonable), the actual timescales should not
be taken seriously. Rather it is their hierarchy the figure is meant to
illustrate. One sees from the left panel that the first thing that happens is
that the planet's spin very quickly approaches the pseudo-synchronous spin rate
(on a timescale of tens of thousands of years in this particular case).
Subsequently, on a roughly 4 orders of magnitude longer timescale, the orbit is
circularized. Finally, on a timescale another one or two orders of magnitude
longer, stellar tides cause the orbit to decay and also reverse the trend of the
stellar spin evolution from spinning down due to magnetic braking (like Sun-like
stars are known to do) to spinning up due to the tidal friction. With the
parameters chosen for this calculation, the planet gets too close to the star
and is destroyed shortly before the end of the simulation.

\section{Equilibrium states}
\label{sec:equilibrium}

\subsection{The end state of tidal evolution}

Let us begin with a significantly simplified problem of tidal evolution. First,
consider a system consisting of just one planet and one star. Second, let us
assume that tides are allowed unlimited time to act. Third, let us ignore all
other processes that can change the system. Under these conditions, one of two
possible final states will eventually be reached. Either the two objects will
merge, or the system will end up in a state satisfying three simultaneous
properties: the orbit is circular, the planetary and stellar spin axes are
aligned with the orbital angular momentum, and the planet and star spin periods
are both equal to the orbital period.

Which of these two possibilities the system approaches depends on the initial
angular momentum available to the system. Since tides are purely internal
interactions within the system, they can only redistribute angular momentum, but
the total angular momentum vector must remain fixed. Let $M_\star$ and $M_p$ be
the masses of the star and planet, and $I_\star$ and $I_p$ their moments of
inertia.  If the two objects are to avoid merging, and reach a final
synchronized spin in a circular orbit with semimajor axis $a_f$, the final spin
angular velocities of both the star and the planet must be equal to the orbital
angular velocity, which by Kepler's third law is:

\begin{equation}
    \Omega_{orb} = \sqrt{\frac{G (M_\star + M_p)}{a_f^3}}
    \label{eq:orbital_angular_velocity}
\end{equation}

This gives the final angular momentum of the system as:

\begin{equation}
    L
    =
    \left(I_\star + I_p\right) \sqrt{\frac{G (M_\star + M_p)}{a_f^3}}
    +
    M_\star M_p \sqrt{\frac{G a_f}{M_\star + M_p}}
    \label{eq:equilibrium_angmom}
\end{equation}

The first term above is the sum of the spin angular momenta of the planet and
the star, and the second term is the orbital angular momentum.

\begin{figure}[t]
    \centering
    \includegraphics[width=0.5\textwidth]{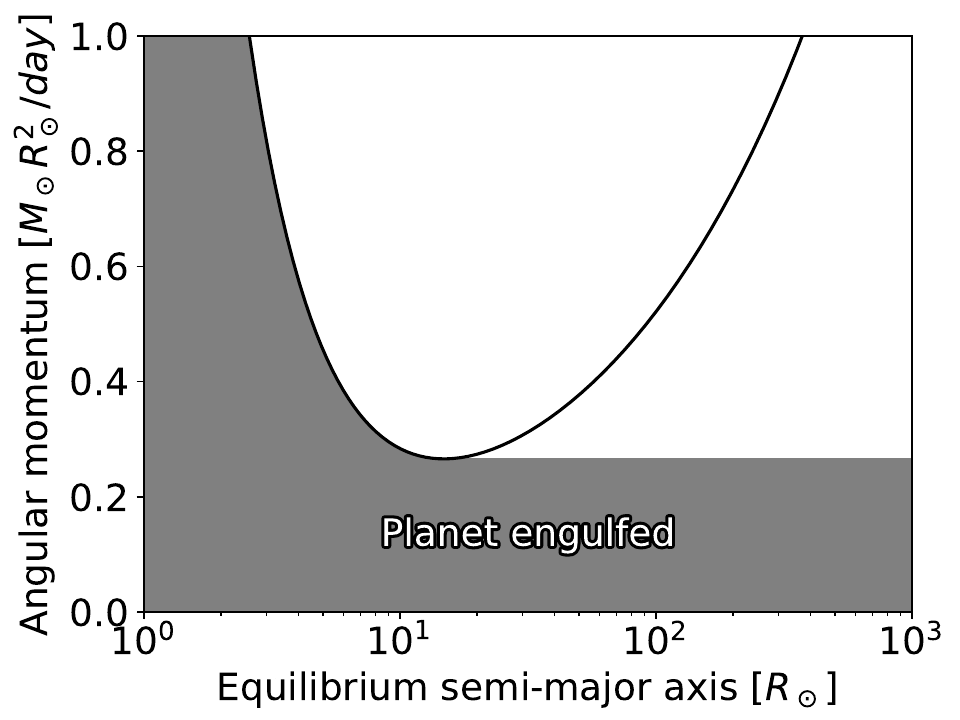}
    \caption{
        The equilibrium angular momentum as a function of the equilibrium
        semimajor axis (Equation \eqref{eq:equilibrium_angmom}) for a
        star-planet system containing a star exactly like the present day Sun
        and a planet exactly like Jupiter.
    }
    \label{fig:equilibrium_angmom}
\end{figure}

From the above equation we see that $L \rightarrow \infty$ both as $a_f
\rightarrow 0$, and as $a_f \rightarrow \infty$, and that $L$ has a minimum
at a finite value of $a_f$. Figure \ref{fig:equilibrium_angmom} shows a graph of
equation \eqref{eq:equilibrium_angmom} for a Jupiter mass planet around a Solar
mass star. The minimum angular momentum for which equation
\eqref{eq:equilibrium_angmom} can be satisfied defines a critical value.  If the
initial angular momentum is smaller than this critical value, no equilibrium
state exists with the two objects surviving, hence the ultimate fate of a system
starting below this critical angular momentum is for the planet to be engulfed
by the star. If the initial angular momentum is larger than the minimum angular
momentum required by the final equilibrium state, then Eq.
\eqref{eq:equilibrium_angmom} allows us to calculate the final semimajor axis
and in turn the spin periods of the planet and the star.

Evidently, if the angular momentum exceeds the critical value, there are two
possible equilibrium states (there are two values of $a_f$ that satisfy Eq.
\eqref{eq:equilibrium_angmom}). From Equations
\eqref{eq:orbital_angular_velocity} and \eqref{eq:equilibrium_angmom} above one
can show that the solution with the smaller semimajor axis is an unstable
equilibrium, while the solution with the larger semimajor axis is in a stable
equilibrium. Physically, if an infinitesimal amount of angular momentum is
transferred from the orbit to the spin of the star (we can ignore the spin
angular momentum of the planet), both the spin and orbital angular velocities
will increase.  However, if a system starts with a semimajor axis equal to the
smaller of the two solutions, the orbital angular velocity will increase more
than the spin, which in turn will ensure tides will transfer even more angular
momentum from the orbit to the spin. If instead the system begins with the
larger semi-major axis solution, the orbital angular velocity will increase less
than the spin angular velocity and tides will work to undo the angular momentum
transfer.  Similar logic applies if one assumes an initial perturbation
transferring angular momentum from the spin to the orbit. In that case both the
orbital and spin angular velocities will decrease, but for the smaller semimajor
axis solution the orbital angular velocity will decrease faster, causing even
more angular momentum transfer from the spin to the orbit. As a result, if the
initial semi-major axis is larger than the smaller of the two solutions, tides
will drive the system toward the larger $a_f$, slower spinning, solution. If the
systems starts interior to the smaller of the two solutions, tides will cause
orbital decay until the planet is destroyed.

In practice, the simplifying assumptions we made above are often violated. Many
exoplanet systems have multiple planets, or even multiple stars. These extra
objects can perturb the orbit, maintaining non-zero eccentricity in spite of
tides. A particularly common situation that arises involves a pair or more
planets in what are called low order mean motion resonances. This mouthful of a
term refers to the situation where a small integer times the orbital period of
one planet is very close to another small integer times the orbital period of
another planet. An example of such mean motion resonance in our own solar system
are three of the four major satellites of Jupiter, where the orbital periods of
Io, Europa, and Ganymede are in a 1:2:4 ratio. Because planets in such
configurations repeatedly get closest to each other in the same place in their
orbit, the gravitational interactions between such planets are particularly
effective in exciting their orbital eccentricities. This excitation can compete
with tidal circularization, maintaining non-zero eccentricity. This prevents
tides from achieving the above equilibrium state. Instead, it can be shown
mathematically that in many situations, as tidal dissipation removes energy from
the systems, the resonant chain of planets will be maintained, causing all
planets to migrate inward together. This could cause the eventual engulfment of
some of the planets by the star even if the initial angular momentum is
sufficient according to the above criterion. In our own solar system, this
common migration seems to occur for Io, Europa, and Ganymede. Except in this
case the system of satellites moves outward rather than inward, because
Jupiter's spin period is shorter than even the shortest period orbit (that of
Io).

Even for a system of just a single planet and a single star, the assumption that
angular momentum is conserved may not hold. In particular, Sun-like stars
continuously lose angular momentum throughout their lifetime. This loss of
angular momentum gets larger the faster the star spins. Tides are generally only
important for orbital periods of order few days or less. In that case, the
equilibrium state described above will correspond to very fast stellar spin, ten
or more times faster than the present day spin of the Sun. At these spin rates,
the loss of angular momentum is significant enough to keep the orbit shrinking,
even if tides keep the system in a circular orbit and the stellar spin
synchronized with the orbit. This should not occur for higher mass stars, since
they appear to not experience appreciable angular momentum loss.

Finally, tides do not have unlimited time to act. Stars have finite lifetimes,
so even for an isolated system with just two objects, the equilibrium state may
not be reached before the star runs out of nuclear fuel and stars expanding.
This may result in the demise of the planet much sooner than due to tides alone.

\subsection{Spin-orbit locking in eccentric systems}

On their way to the synchronized and circularized state described above, planets
may find themselves crossing orbital configurations where the rate of tidal
evolution slows down dramatically. These quasi-equilibrium states require a
significantly eccentric orbit. For circular orbits, the time dependence of the
tidal forces experienced by any point within the planet or the star are simple
sine waves with a fixed frequency equal to twice the difference between the spin
and orbital angular velocity of the corresponding object. As already discussed
above, the direction of tidal evolution switches as this frequency crosses zero.

For eccentric orbits, the time dependence of the tidal forcing is more
complicated. Let us break up the problem into two parts. First, consider a
non-rotating object in an eccentric orbit. As already discussed above, we can
calculate the average effect of tides over a single orbit, ignoring the changes
in the orbit or the spin of the objects. Under this approximation, the tidal
forcing experienced by any part of a non-spinning object will be periodic with
the orbital period. As a result, we can think of the tidal forcing as a discrete
Fourier series of multiple tidal waves, with frequencies $m' \Omega_{orb}$,
where $\Omega_{orb}$ is the orbital frequency and $m'$ is an integer. If we now
include the rotation of the object with angular velocity $\Omega_{spin}$ and
assume its axis of rotation is aligned with the orbital angular momentum, each
of these waves will have its frequency shifted by $2\Omega_{spin}$, giving us a
series of tidal waves with frequencies $\Omega_{tide} = m' \Omega_{orb} - 2
\Omega_{spin}$. The factor of two comes from the fact that there are two tidal
bulges, one on the side facing the companion object and one on the opposite
side. In principle, for non-circular orbits, tidal waves exist for arbitrarily
large $m'$, though their amplitude becomes negligible as $m'\rightarrow\infty$.
Even more generally, if we now allow the axis of rotation to have any
orientation relative to the orbit, we end up with an even more general series of
tidal waves with frequencies:

\begin{equation}
    \Omega_{tide} = m' \Omega_{orb} - m \Omega_{spin},
    \quad m' \in \mathbb{Z},\quad m \in \left\{0, 1, 2\right\}
\end{equation}

At this point we need to introduce another simplifying assumption. Namely, we
will assume that the tidal perturbations are weak enough so we do not need to
worry about interactions between the separate tidal waves and instead we can
calculate the tidal torque and power each wave applies to the orbit separately
and calculate the evolution by simply adding these up.

Just like for the single tidal term of aligned circular orbits, the tidal torque
from each of these tidal terms changes its sign as the frequency of that term
crosses zero. If the amplitude of the tidal term for a given $m,m'$ is large
enough and the increase in dissipation as the frequency moves away from zero is
sufficiently steep, this change in the sign of the tidal torque introduces the
possibility that the spin of the object can be locked in a frequency close to
$\Omega_{spin}=\frac{m'}{m} \Omega_{orb}$. This can happen if the tidal torque
of that particular tidal term is large enough to cancel the combined torques of
all other terms. As the relative amplitudes of the different tidal terms depend
on the eccentricity, the set of frequencies at which the spin of the object can
be locked depends on the eccentricity of the orbit. The closer the orbit is to
circular the smaller the amplitude of the tidal terms with large $m'$, and hence
fewer terms will have sufficient amplitudes to hold the spin in a locked state.

In principle, both the planet and the star spin for eccentric systems can be
locked in one of these states. In our own Solar system, Mercury is in such a
spin-orbit locked state, with $m'=3$ and $m=2$, i.e. executing three revolutions
for every two orbital periods. It is likely that at least some exoplanets in
short orbital periods are also locked in similar states, as long as their orbits
have sufficient eccentricity to support the lock. Unfortunately, so far we have
not found a way to measure the spin of exoplanets, so we do not have a way of
observing this effect directly.

It is important to note that, even though the overall tidal torque is close to
zero if an object is in one of these pseudo-equilibrium states, tides in that
object can still contribute to the evolution of the orbit, since energy will
still be dissipated. The only truly stable tidal configuration is a circular
orbit with the spins of both objects synchronized with the orbit and their
equatorial planes coinciding with the orbital plane. And as pointed out above,
in the presence of non-tidal processes, even that may not be a configuration
which can last indefinitely.

\section{Observational evidence for tidal evolution}
\label{sec:observables}

The effect of tides scales as the difference in the gravitational acceleration
due to the companion at the near and far tidal bulges multiplied by the mass in
each bulge. Both of these factors themselves are a strong function of the ratio
of the size of the object to the size of the orbit. As a result, the tidal
coupling between a planet and its host star decreases very rapidly as the
distance between the planet and the star increases. This is why tides are only
important for planets which get very close to their parent stars at least for
some part of their orbit. So a universal property of tides is that they should
leave observable footprints only in the systems with the smallest star-planet
separations, corresponding to the shortest orbital periods and that all effects
will diminish at longer orbital periods.

As described in the previous section, planetary tides exchange angular momentum
between the orbit and spin of the planet, stellar tides exchange angular
momentum between the orbit and spin of the star, and both tides extract energy
out of the system, leading to orbital circularization.  To reiterate what we
already mentioned in Section \ref{sec:timescales}, the quantity that is most
readily affected by tides is the spin of the planet. This is due to two reasons.
First, the angular momentum of the planet is many orders of magnitude smaller
than both the orbital and stellar angular momenta, so it most easily changed.
Second, the self-gravity of the planet is much smaller than that of the star,
and the tidal force on the planet is much larger than that on the star. Since
the size of the tidal bulges is determined by the competition between
self-gravity and tidal force, the planetary tides are much stronger. As a
consequence, the expectation is that pseudo-synchronizing the planet's spin with
the orbit should be the tidal signature affecting the largest number of
exoplanet systems. Unfortunately, at present, there is no way to observationally
determine the spin of an exoplanet, so we cannot test this prediction.

\subsection{Tidal circularization}
\label{sec:circularization}

\begin{figure}[t]
    \centering
    \includegraphics[width=0.5\textwidth]{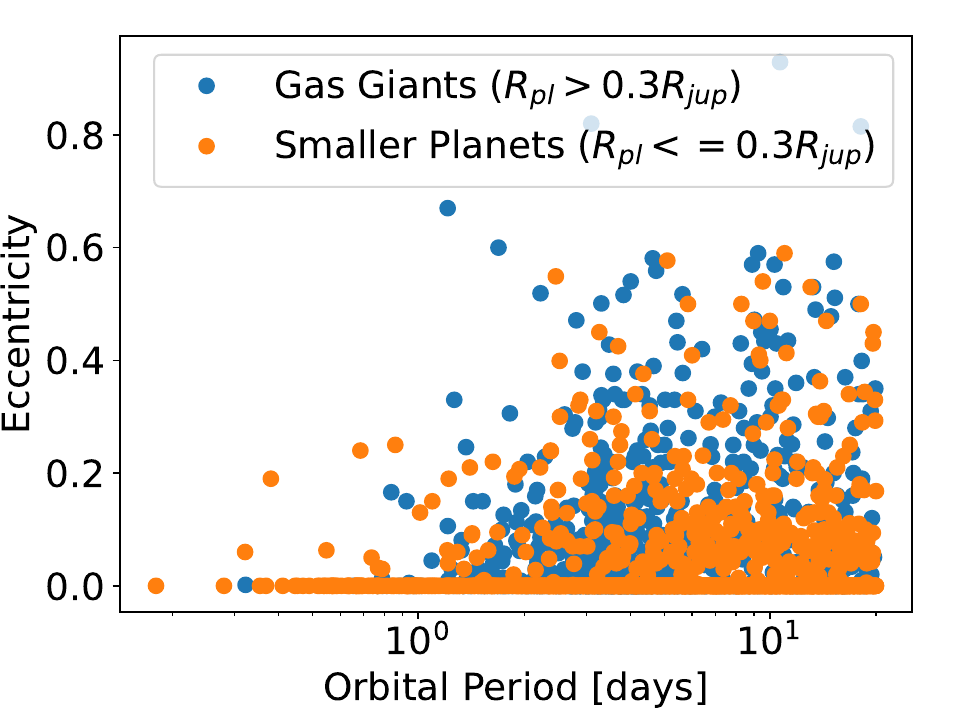}
    \caption{
        The period vs eccentricity of the currently confirmed extrasolar planets
        from the NASA exoplanet archive. The color indicates the radius of the
        planet.
    }
    \label{fig:period-eccentricity}
\end{figure}

Based on the timescale hierarchy of Section \ref{sec:timescales}, the next most
strongly affected property of exoplanet systems is the orbital eccentricity.
For most systems the orbital angular momentum is comparable to or smaller than
the stellar spin angular momentum and it is feeling the combined effect of both
stellar and planetary tides, with the latter being significantly more important
as long as the eccentricity is not negligible. There is well established
observational evidence that tides are affecting the orbital eccentricities of
exoplanets. Namely, the shortest period systems, which experience the strongest
tides and hence should get circularized very quickly, are almost all observed in
circular orbits, with gradually more and more eccentricity surviving at longer
and longer orbital periods (see Fig.~\ref{fig:period-eccentricity}).

There is an important difference between circularization due to stellar vs.\
planetary tides. Each type of tide couples the orbit to the spin of the
corresponding object. Since the planet's spin angular momentum is negligible
compared to the orbital angular momentum, to an excellent approximation,
planetary tides cannot change the orbital angular momentum. This can be shown to
imply that if only planetary tides are present, circularization will proceed in
such a way as to hold $r_p(1+e)$ constant, where $r_p$ is the so called
pericenter distance (the distance of closest approach between the planet and the
star) and $e$ is the eccentricity. In other words, under only planetary tides,
circularization is inevitably accompanied by an increase in the distance of
closest approach between the star and the planet. Since tides are strongest at
pericenter, this generally decreases the rate of circularization over time.

This is in contrast to the case of stellar tides, where the stellar spin angular
momentum is comparable to or perhaps even dominates over the orbital angular
momentum. In this case, the pericenter distance can grow or shrink, depending on
the stellar spin and orbital configuration. Sun-like stars in particular have
spin periods longer than a few days for most of their main-sequence lifetime.
Since tidal circularization is only important for orbital periods of a few days
or less, stellar tides will mostly decrease the angular momentum of the orbit
over time, resulting in smaller pericenter distance compared to planetary
tides, and thus less pronounced decrease in the rate of circularization as the
orbit circularizes.

\subsection{Tidal inspiral}

Another effect that is expected to occur in very short period exoplanet systems
is tidal inspiral. Based on Section \ref{sec:timescales}, this is only relevant
for exoplanet systems whose orbital period is short enough for the tides on the
planet to have circularized the orbit and synchronized the spin of the planet
with it. As a result, the planetary tides are now static on the surface of the
planet and no longer experience friction. This leaves only stellar tides to
drive further evolution.  Stars generally have spin periods exceeding the
orbital period. As discussed above, this situation leads to tidal bulges that
lag behind the planet as it move around in its orbit, leading to removal of
angular momentum from the orbit and adding it to the spin of the star. Since
stars have hundreds to thousands of times more mass than even the most massive
planets, their moment of inertia is in many cases large enough to prevent the
star from synchronizing to the orbital period of the planet. That is, the
angular momentum of systems where tides are important is often below the
critical value defined by Eq.  \ref{eq:equilibrium_angmom}. As a result, stellar
tides gradually shrink the orbit of the planet, driving it closer and closer to
the star.

If the planet starts out relatively far away from the star, this inspiral could
take longer than the lifetime of the star. However, for the shortest period
systems, as the planet gets closer to the star tides get stronger and the
inspiral accelerates. Thus a positive feedback loop is established. This will
eventually push the planet close enough to its parent star for one of two things
to happen. First, the strengthening of the planetary tides as the planet-star
separation decreases could eventually result in a tidal force attempting to
stretch the planet that overwhelms the self-gravity of the planet. This will
cause the planet to be tidally ripped apart. The distance between the planet and
the star at which this occurs is known as the Roche limit, and it depends on the
size and mass of the planet. The more massive the planet is, the stronger its
self gravity, pushing the Roche limit closer to the star.  Similarly, the larger
the planet, the weaker its self gravity, and hence the Roche limit is pushed
further away. In the case of gas giant planets, the Roche limit is comparable to
the size of the star and for rocky planets it is considerably smaller. For
planets for which the Roche limit lies below the stellar surface, the planet
will be engulfed by the star and destroyed that way. In either case, the tidal
inspiral will cause the planet to be lost.

Currently, a number of researchers have claimed to directly detect the shrinking
of the orbits of exoplanets over time by observing a shortening of the time
between consecutive transits. For example, the orbit of the very first planet
discovered by the \textit{Kepler} space satellite Kepler-1658 (a.k.a KOI-4)
appears to be decaying \citep{Vissapragada_et_al_22}.
\citet{Maciejewski_et_al_16} also claim to detect orbital decay in the system
WASP-12. This finding was independently confirmed by \citet{Patra_et_al_17}. At
the moment it is hard to rule out alternative explanations for these
observations. In most cases it is still possible to reproduce the observed
shifts in transit times by invoking mechanisms others than orbital decay. For
example orbital precession or the effect of undetected third bodies in these
exoplanet systems can still reproduce the observations. In some cases these
alternatives are disfavoured, but to date they have not been completely ruled
out for any particular system.

Two separate indirect lines of evidence point to this process occurring for gas
giant exoplanets, while perhaps not for smaller, rocky planets. First,
\citet{Jackson_et_al_09} point out that the smallest orbits in which younger
exoplanet systems are observed are smaller than the smallest orbits in which
older systems are observed. This matches the predictions of the tidal inspiral
scenario. Since older systems have more time to be affected by tides and since
tidal decay accelerates very quickly as the orbit gets smaller, over time tides
will clear out progressively longer period systems.

A different line of evidence for the tidal destruction of the shortest orbital
period exoplanets comes from their velocities within the galaxy.
\citet{Hamer_Schlaufman_19} compare the host stars of giant planets in orbits
with periods few days (a.k.a.  hot Jupiters) to two control samples of stars
without planets, and stars with longer period planets. What they find is that
the velocities of the hot Jupiter hosts show much smaller scatter around the
mean rotation of the galaxy compared to either of the control samples, while the
distribution of velocities of the two control samples are statistically
indistinguishable. It is a well established property of our galaxy that the
smaller the scatter in the velocities of a sample of stars, the younger it is.
Hence, this observation is another indication of the relative youth of systems
containing short period giant planets, suggesting the older systems have lost
their planets to tidal inspiral.

\subsection{Tidal spin-up of Sun-like exoplanet host stars}

Sun-like stars gradually spin down over time. An important property of this spin
down is that faster spinning stars lose angular momentum faster. As a result,
even if stars start with a brad distribution of initial spins, over time the
distribution narrows, causing isolated Sun-like stars older than a few hundred
million years to have a spin period tightly related to their age. Measuring the
rotation and using this relationship is one of very few methods of determining
the ages of isolated stars.

For stars experiencing tides however, this unique spin-age relationship can be
broken. For exoplanets experiencing tidal inspiral the host star is expected to
spin faster than an isolated star of the same age would. This is direct
consequence of angular momentum conservation. As tides drive the planet closer
to its parent star, the orbital angular momentum that is lost gets added to the
stellar spin. One would expect then that stars with short period giant
exoplanets orbiting them will spin on average faster than similar isolated stars
or stars with longer period or small planets. In fact, \citet{Tajeda_et_al_21}
claim to confirm this prediction.

On the one hand, this is unfortunate, since measuring the rotation period of
stars and using the spin-age relationship is one of very few ways of determining
stellar ages. The fact that tides affect the spin, means this method can not be
reliably used for hot Jupiter hosts. On the other hand, \citet{Penev_et_al_18}
suggest that the deviation from single star spin evolution is significant enough
to be used as a probe of the tidal dissipation physics of Sun-like stars. The
latter is particularly important as exoplanet systems are sensitive to a
different regime of tides than binary stars. For systems with measurable tidal
effects, the tides a planet raises on its parent star systems are simultaneously
smaller amplitude and shorter period than the tides raised by the two components
of a binary star on each other. Since only stellar tides affect the stellar
spin, unlike the effects on the orbit for which both stellar and the planetary
tides are important, observing the tidal spin-up of exoplanet host stars allows
probing the tidal physics of stars in this new regime.

\subsection{Tidal alignment}

The tidal torque due to the stellar tides on the orbit will also cause the orbit
to align with the equator of the star. This is one of the leading explanations
for the observed trend that giant planets orbiting Sun-like stars appear to have
orbits that are aligned with the stellar equator, while stars significantly more
massive, and hence hotter, than the sun, seem to host planets with all kinds of
orbital orientations \citep[c.f. Fig 3 and Fig 8 of][]{Albrecht_et_al_22}. The
lack of alignment between hot stars and the orbits of their companion planets
has been hypothesized to be due to much weaker tidal friction that is expected
to occur in these stars. More specifically, the transition between aligned and
misaligned orbits appears to coincide with the transition between stars that
have surface convective zones and those that do not, and the turbulent flow
present in stellar convective zones has long been thought to be one of the
leading causes of tidal friction.

It must be emphasized, however, that the tidal interpretation of the observed
alignment pattern is far from universally accepted.  Alternative suggestions
include the possibility that most planets generally form aligned, with various
mechanisms causing misalignment in some systems. That said, the tidal
explanation is one of the most compelling, since it does not require assuming
any new physics. It also appears to be supported by the exceptions observed by
\citet{Winn_Fabrycky_2015}, namely that significant misalignments of planets
around Sun-like stars seem to occur for planets orbiting at wider orbits or for
smaller planets, both of which correspond to weaker tides.

In general, all proposed explanations for the observed alignment pattern,
including the tidal one, have at least one significant problem they need to
overcome. In the case of tides, the issue is one of timescales. As we discussed
in the introduction, the timescale for tidal inspiral is comparable to or
shorter than the timescale on which the stellar spin is affected by tides. As a
result, any system which has been appreciably tidally aligned, should also have
experienced significant inspiral. Because all tidal timescales decrease very
dramatically as the orbit gets smaller, this in turn would imply that the planet
has only a short time remaining to live before tides drive it into the star.
This simple reasoning suggests that if the observed obliquity pattern is
explained by a simple tidal model, most observed planets would be very near the
end of their lifetime. This is highly statistically implausible. Instead one
would expect planets to be distributed more uniformly with age. While detecting
shorter period planets is somewhat easier than longer period ones, the
difference is not large enough to match what would be naively predicted if tides
are responsible for the observed alignment pattern. \citet{Lai_12} and
\citet{Anderson_et_al_21} suggest that this problem could be resolved by more
realistic tidal dissipation theories than the somewhat simplistic treatment
discussed in the introduction. Both publications invoke tidal dissipation that
gets weaker for tidal waves with high frequencies, though in a very different
manner. Since tidal alignment is driven by significantly lower frequency tidal
waves than orbital decay, this decouples the two timescales and allows for a
more statistically acceptable distribution of exoplanet systems over their
lifetimes.

\subsection{Tidally heating and inflating exoplanets}

Tidal friction converts some of the mechanical energy of tidal deformations into
heat. This heating is irrelevant for the star, since it only constitutes a
negligible fraction of the stellar luminosity. However, for the planet this may
not be the case.

In our own solar system we have several rather dramatic examples where tidal
heating drives spectacular geological processes. Two dramatic examples are
Jupiter's moon Io and Saturn's moon Enceladus. In particular, Io is perhaps the
most geologically active body in the Solar system even though it has a similar
size to the Moon and hence should cool on a similar timescale. The heating
required to maintain that activity is attributed to tidal friction. Similarly,
tidal heating of Enceladus is thought to be responsible for powering spectacular
eruptions of water ice particles and gas from the moon's south pole.  The best
to-date accounting for the energy budget of Enceladus suggests that this
geologically active region on Enceladus accounts for roughly a quarter of the
entire energy budget of that moon, even though it covers less than 5\% of the
surface.

In the case of exoplanets, we do not yet have any firm direct evidence of the
importance of tidal heating, but it has been speculated to be an important
mechanism affecting the sizes of hot Jupiters. Many hot Jupiters are observed to
have sizes that are larger than can be explained by standard models of planetary
structure, and one of the proposed explanations for these inflated radii is
tidal heating.

Even though the energy delivered by tidal dissipation may be small compared to
direct irradiation by the star, it can have an outsized impact. Because, even
though stellar irradiation deposits massive amounts of energy, it does so at
optical depths of order unity, where it is quickly radiated away without
significantly affecting the internal structure of the planet. In contrast,
\citet{Komacek_Youdin_17} argue that even orders of magnitude smaller heating,
if deposited deeper into the planet can have a profound effect on the planetary
radius.

In order for tides to heat, and as a result, inflate a planet, the planet needs
to either be in an orbit with non-zero eccentricity, or have a spin period that
is not exactly equal to its orbital period. For planets spinning synchronously
in circular orbits the tidal deformation is static, and no energy is dissipated.
Thus, tidal heating can be a temporary phenomenon, typically early in the
lifetime of a planet. For example, if a planet begins in a significantly
eccentric orbit it  may experience significant tidal heating until its orbit is
circularized. Such situations will typically occur early in the life of an
exoplanet system. In fact, excitation of the eccentricity to very large values,
followed by tidal circularization is one of the mechanisms proposed for how some
of the shorter period exoplanets come to be in their current orbits. However, by
the time the host stars reach the main sequence, such processes will mostly have
played out, leaving planets experiencing strong tides in circular orbits with
their spins synchronized to the orbital period.

Alternatively, in realistic systems, it is possible to maintain non-zero
eccentricity long-term for example if it is excited by gravitational
interactions with other planets or stars in the system. Since our planet
detection methods are not 100\% effective, even if no additional planets are
observed they may be there.  Particularly important are planets residing close
to mean motion resonances as described in Section \ref{sec:equilibrium}. In the
case of Enceladus mentioned above, the orbit is observed to be slightly
eccentric. Its orbital eccentricity is maintained by interactions with one of
the other moons of Saturn: Dione, which is in a 2:1 mean motion resonance with
Enceladus.

It is also possible that some mechanism counteracts tidal synchronization, again
maintaining some degree of asynchrony.  One such proposed mechanism are the
so-called thermal (or atmospheric) tides.  Very briefly, the term thermal tides
refers to fluid motions in the atmosphere of the planet that are driven by
time-variable stellar irradiation. Variable irradiation can arise because of
orbital eccentricity, a significant tilt of the planet's axis of rotation
relative to the orbital plane (e.g. the seasons on Earth), or because of
asynchronous rotation.  The fluid mechanical response to such variable heating
of the atmosphere is in general quite a complicated process that is outside the
scope of this introductory chapter, but the ultimate result is that the
resulting redistribution of mass can alter the tidal potential of the planet in
addition to the effects of its tidal distortion. In turn this modified tidal
potential results in the equilibrium eccentricity of the orbit being non-zero
and/or the equilibrium spin of the planet to not be precisely synchronous. If
these departures from synchrony and circularity are large enough, tidal
dissipation within the planet can produce significant heat, possibly inflating
the planet, and accelerating its orbital decay beyond what the tidal dissipation
within the star alone can provide. \citet{Arras_Socrates_10} argue that at least
in principle, this effect could provide sufficient heating to explain the
observed radius inflation of hot Jupiters.

\section{Conclusions}
Research aiming to understand and model tidal interactions has entered a period
of renewed interest. This is mostly driven by the discovery of thousands of
exoplanets in short orbital periods which experience strong star-planet
interactions. Serendipitously, the same space- and ground-based surveys that
detect exoplanets also provide vast amounts of new data which can be used to
probe tidal friction and other tidal processes. This includes the exoplanet
systems themselves as described in this chapter, but also a wealth of binary
star systems in short period orbits which are characterized with orders of
magnitude higher precision than was previously available. Furthermore, the
increasing availability of computing resources allows ever more sophisticated
direct simulations of aspects of tidal interactions under ever more realistic
conditions. Thus the extremely active field of exoplanets simultaneously drives
interest and provides new data to study tidal interactions and this couples very
favorably with increased computing capacity to suggest that the renewed interest
in tides will continue and possibly intensify in the near future.

\bibliographystyle{Harvard}
\bibliography{bibliography}

\end{document}